\documentclass[12pt,epsfig,prd,showpacs]{revtex4}
\usepackage{graphicx}
\usepackage{epsfig}
\usepackage{dcolumn}
\usepackage{bm}

\newcommand{\etap}{\eta^{\prime}}
\newcommand{\jpsi}{J/\psi}

\newcommand{\pppm}{\pi^+\pi^-}
\newcommand{\ar}{\rightarrow}

\begin{document}
\vspace{-30mm}
\title{\bf \boldmath Measurement of the branching fractions for
$J/\psi\rightarrow\gamma\pi^0$, $\gamma\eta$ and $\gamma\eta^{\prime}$}
\author{
M.~Ablikim$^{1}$,              J.~Z.~Bai$^{1}$,               Y.~Ban$^{11}$,
J.~G.~Bian$^{1}$,              X.~Cai$^{1}$,                  H.~F.~Chen$^{16}$,
H.~S.~Chen$^{1}$,              H.~X.~Chen$^{1}$,              J.~C.~Chen$^{1}$,
Jin~Chen$^{1}$
% ...
\\(BES Collaboration)\\
\vspace{0.2cm}
$^{1}$ {\it Institute of High Energy Physics, Beijing 100049, People's Republic of China}\\
% ...
$^{11}$ {\it Peking University, Beijing 100871, People's Republic of China}\\
% ...
$^{16}$ {\it University of Science and Technology of China, Hefei 230026, People's Republic of China}\\
% ...
}
\begin{abstract}
 The decay modes
 $\jpsi\ar\gamma\pi^0, \gamma\eta$ and $\gamma\etap$ are analyzed using 
 a data sample of 58 million $\jpsi$ decays 
 collected with the BESII detector at  BEPC. The branching fractions are
 determined to be: 
 $Br(\jpsi\ar\gamma\pi^0)=(3.13^{+0.65}_{-0.44})\times10^{-5}$, 
 $Br(\jpsi\ar\gamma\eta)=(11.23\pm0.89)\times10^{-4}$, and 
 $Br(\jpsi\ar\gamma\etap)=(5.55\pm0.44)\times10^{-3}$, where the errors are combined 
 statistical and systematic errors. The ratio of partial widths 
 $\Gamma(\jpsi\ar\gamma\etap)/\Gamma(\jpsi\ar\gamma\eta)$ is measured to be $4.94\pm0.40$, and the 
 singlet-octet pseudoscalar mixing angle of $\eta-\etap$ system is determined to be 
 $\theta_{P}=(-22.08\pm0.81)^{\circ}$.

\end{abstract}
\maketitle


\vspace{5mm}
\section{INTRODUCTION}

 In flavor-$SU(3)$, the $\pi^0$, $\eta$ and $\etap$ mesons 
 belong to the same pseudoscalar nonet. 
 The physical states $\eta$ and $\etap$ are related to the $SU_f(3)$-octet state $\eta_8$ and the 
 $SU_f(3)$-singlet state $\eta_1$, via the usual mixing formulae:
$$
\eta=\eta_8\cos\theta_P-\eta_1\sin\theta_P,
$$
$$
\etap=\eta_8\sin\theta_P+\eta_1\cos\theta_P,
$$
where $\theta_P$ is the pseudoscalar mixing angle~\cite{theo,mixing}. 
The conventional estimate of $\eta - \etap$ mixing uses the quadratic mass matrix
\[
M^2 = \left(\begin{array}{cc}  M^2_{88} & M^2_{18} \\ M^2_{18} & M^2_{11}\end{array}\right),
\]
where  $M^2_{88}=\frac{1}{3}(4m^2_K-m^2_{\pi})$ is given by the Gell-Mann-Okubo mass formula. 
Diagonalization of this matrix gives
$$
\tan^2 \theta_P = \frac{M^2_{88}-m^2_{\eta}}{m^2_{\etap}-M^2_{88}} \Longrightarrow \theta_P \approx -10^{\circ}.
$$
With a linear mass matrix and the linear 
Gell-Mann-Okubo mass formula  $M_{88}=\frac{1}{3}(4m_K-m_{\pi})$, 
$\theta_P$ is computed to be about $-24^{\circ}$~\cite{mixing}. 

The mixing angle has been measured experimentally in different ways, and the value is around 
$-20^{\circ}$~\cite{mixing}. One of these measurements is based on $\jpsi$ radiative decays. 
In the limit where the OZI rule and $SU_f(3)$ symmetry are exact, one gets~\cite{thetap1}
$$
R =
 \frac{\Gamma(\jpsi\ar\gamma\etap)}
 {\Gamma(\jpsi\ar\gamma\eta)}=(\frac{p_{\etap}}{p_{\eta}})^3\cdot{\cot^2\theta_P},
$$
where $p_{\eta}$ and $p_{\etap}$ are the momenta of $\eta$ and $\etap$
in the $\jpsi$ Center of Mass System (CMS). 

The first-order perturbation theory~\cite{pertu1,pertu2} 
expression for the partial width  $\Gamma(\jpsi\ar\gamma + 
 pseudoscalar)$ is 
$$
\Gamma(\jpsi\ar\gamma + P) = \frac{1}{6}(\frac{2}{3})^2\alpha^4_s \alpha Q^2_c\frac{1}{M^3_{\jpsi}}(\frac{4R_{\jpsi}(0)}{\sqrt{4\pi M_{\jpsi}}})^2 (\frac{4R_{P}(0)}{\sqrt{4\pi M_{P}}})^2 x |H^P(x)|^2.
$$
Here $R_{\jpsi}(0)$ and $R_{P}(0)$ are the wave functions at the origin 
of the $\jpsi$ and the pseudoscalar with mass $M_{P}$, and $Q_c$ is the charge of the charmed
quark. The pseudoscalar helicity amplitude $H^P(x)$ depends on $x=1-(\frac{M_P}{M_{\jpsi}})^2$; 
numerically $x|H^P(x)|\approx55$ for $M_P=m_{\etap}$. $R_{\jpsi}(0)$ and $R_{P}(0)$ can be 
determined from the $\jpsi\ar e^+ e^-$ and $P\ar\gamma\gamma$ partial decay widths, 
respectively. 
Using the lowest-order QCD formula for $\alpha_s$, the $\jpsi\ar\gamma\etap$ decay width is 
calculated to be 213~eV, which is in agreement with the experimentally measured value. The value of $\Gamma(\jpsi\ar\gamma\eta)$
determined from the same 
formula disagrees with measurements.
Some models that assign a small admixture of $\eta$ and $\etap$ to other states have 
been proposed to explain the large value 
of the ratio $R=\Gamma(\jpsi\ar\gamma\etap)/\Gamma(\jpsi\ar\gamma\eta)$. 
For example, Ref.~\cite{ratio1}, which  assigns small $c \bar{c}$ contribution from $\eta_c$ in the $\eta$ and $\etap$ wave functions, 
predicts $R=3.9$;  Ref.~\cite{ratio2} gives a value of $R=5.1$ by 
considering some admixture of the $\iota(1440)$ to the $\eta$ and $\etap$. A precision 
measurement of $R$ could distinguish between these mixing models, as well as provide
a determination of the mixing angle 
$\theta_P$.  Experimental measurements of $Br(\jpsi\ar\gamma\eta)$ and $Br(\jpsi\ar\gamma\etap)$ 
were reported by the DESY-Heidelberg group~\cite{DESY}, the Crystal Ball~\cite{CBAL}, 
MarkIII~\cite{MARK3} and DM2~\cite{DM2}. 

The decay $\jpsi\ar\gamma\pi^0$ is suppressed because the photon can only be radiated from the 
final state quarks.  This branching fraction was measured by DASP~\cite{DASP} and Crystal 
Ball~\cite{CBAL}; the average of the measurements, $(3.9\pm1.3) \times 10^{-5}$~\cite{PDG04}, 
is in  agreement with the VMD prediction $3.3\times 10^{-5}$~\cite{VMD}. In contrast, the QCD 
multipole 
expansion theory~\cite{mutipole} predicts a value of $1\times10^{-6}$. 

In this paper, $\jpsi\ar\gamma\pi^0$ is studied using $\pi^0\ar\gamma\gamma$ decay, 
$\jpsi\ar\gamma\eta$ is measured using $\eta\ar\gamma\gamma$ and $\eta\ar\pi^0\pi^+\pi^-$ with 
$\pi^0\ar\gamma\gamma$, and $\jpsi\ar\gamma\etap$ is studied using $\etap\ar\gamma\gamma$, 
$\etap\ar\gamma\pi^+\pi^-$ and $\etap\ar\eta\pi^+\pi^-$ with $\eta\ar\gamma\gamma$. The analyses 
use a data sample that contains $58\times10^6$ $\jpsi$ decays collected with the 
updated BEijing Spectrometer 
(BESII) operating at the Beijing Electron Positron Collider (BEPC). 

\section{BES DETECTOR AND MONTE CARLO SIMULATION}

BESII is a large solid-angle magnetic spectrometer that is described in detail in
Ref.~\cite{bes2}. The momentum of  charged particles is measured
in a 40-layer cylindrical main drift chamber (MDC) with a
momentum resolution of $\sigma_{p}$/p=$1.78\%\sqrt{1+p^2}$ ($p$ in
GeV/c).  Particle identification is accomplished using specific
ionization ($dE/dx$) measurements in the drift chamber and
time-of-flight (TOF) information from a barrel-like array of 48
scintillation counters. The $dE/dx$ resolution is
$\sigma_{dE/dx}\simeq8.0\%$; the TOF resolution for Bhabha events is
$\sigma_{TOF}= 180$ ps.  Radially outside of the time-of-flight
counters is a 12-radiation-length barrel shower counter (BSC)
comprised of gas tubes interleaved with lead sheets. The
BSC measures the energy and direction of photons with resolutions of
$\sigma_{E}/E\simeq21\%\sqrt{E}$ ($E$ in GeV), $\sigma_{\phi}=7.9$
mrad, and $\sigma_{z}=2.3$ cm. The iron flux-return of the magnet is
instrumented with three double layers of proportional counters
that are used to identify muons.

A GEANT3-based Monte Carlo simulation package~\cite{simbes}, which simulates the
detector response including interactions of secondary particles in
the detector material, is used to determine detection efficiencies
and mass resolutions,  optimize
selection criteria, and estimate backgrounds.  Reasonable agreement
between data and MC simulation is observed for various calibration
channels, including $e^+e^-\to(\gamma) e^+ e^-$,
$e^+e^-\to(\gamma)\mu^+\mu^-$, $J/\psi\to p\bar{p}$, $J/\psi\to\rho\pi$
and $\psi(2S)\to\pi^+\pi^- J/\psi$, $J/\psi\to l^+ l^-$.

\section{DATA ANALYSIS}

\subsection{\boldmath $\jpsi\ar\gamma\gamma\gamma$}
In the $\jpsi\ar\gamma\gamma\gamma$ decay mode, there are no charged tracks in the final states.
Each candidate event is required to have three and only three photon candidates; the MC 
indicates that the number of these decays that produce final states with
more than three photon candidates is negligible.
 A photon candidate is defined as
a cluster in the BSC with an energy deposit of more than 50 MeV, and with an angle 
between the development direction of the cluster and the direction 
from the interaction point to the first hit layer of the BSC that is 
 less than 20$^{\circ}$. If two clusters have an opening angle 
that is less than 10$^{\circ}$ or have an 
 invariant mass that is less than 50 MeV$/c^2$, the lower energy cluster is regarded as a 
remnant  from the other and not a separate
photon candidate. A kinematic fit that conserves energy and momentum  is 
 applied to the three photon candidates, and $\chi^2\leq20$ is required. We also
require $|\cos\theta_v|<0.8$ and   $\theta_{min}>6^{\circ}$, 
where $\theta_v$ is the polar angle of a decay photon in the pseudoscalar's CMS 
 (shown in Fig. 1a), and $\theta_{min}$ is the minimum angle between any two of the three photon 
 candidates (shown in Fig. 1b).  This rejects background from the
continuum  $e^+e^-\ar\gamma\gamma(\gamma)$ process.

\begin{figure}[htbp]
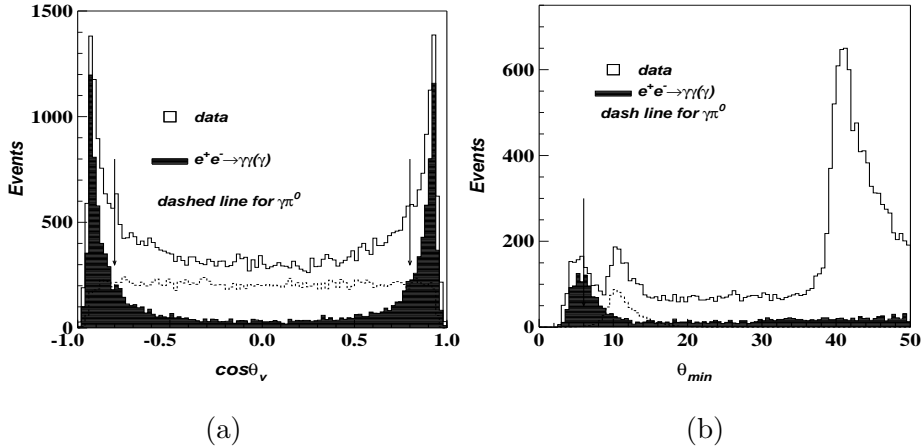

\centerline{\psfig{file=1a.epsi,width=6cm,height=5cm}
            \psfig{file=1b.epsi,width=6cm,height=5cm}}
\parbox[top]{8cm}{(a)~~~~~~~~~~~~~~~~~~~~~~~~~~~~~~~~~~~~~~~~~~~~~~(b)}
\caption{Distribution of (a) $\cos\theta_v$ and (b) $\theta_{min}$. The open histograms are 
$\jpsi$ data, the shaded histograms are background from $e^+e^-\ar\gamma\gamma(\gamma)$, and the 
dashed lines are simulated $\jpsi\ar\gamma\pi^0\ar\gamma\gamma\gamma$ events (not normalized).}
\end{figure}

\subsubsection{$\jpsi\ar\gamma\pi^0,\pi^0\ar\gamma\gamma$}
 
Figure 2 shows the invariant mass distribution in the $\pi^0$ mass region
of the two photon candidates that have the smallest opening 
angle.  A peak at the $\pi^0$  mass is evident.

\begin{figure}[htbp]
\centerline{\psfig{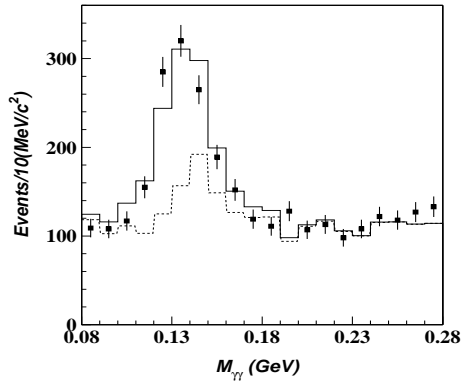}}
\caption{Invariant mass distribution of the $\gamma\gamma$ with smallest opening angle for 
$\jpsi\ar\gamma\gamma\gamma$ candidate events. The solid squares with error bars are data, the 
histogram is the best fit described in the text, and the dashed line is the background.}
\end{figure}

From  MC studies, background channels that produce a peak in the 
$\pi^0$ signal region come mainly from channels with $5\gamma$  
final states, such as $\jpsi\ar\gamma\pi^0\pi^0$, via the $f_2(1270)$, $f_0(2100)$ etc. 
($\jpsi\ar 4 \gamma$s violates C-parity).  These background sources are
studied using events where the number of photon candidates in the event is four. 
Four photon events are  selected and subjected to a four-constraint kinematic fit to 
$\jpsi\ar\gamma\gamma\gamma$,  using any three of the four photons;  the three-photon 
combination with the smallest $\chi^2$ is selected for the background study. Figure~3 
shows the  invariant mass 
distribution for the two photons with the smallest opening angle from four-photon 
events.  A peak is observed in the $\pi^0$ mass region that agrees with 
expectations from MC simulations that include all known modes that produce 
$5\gamma$ final states. 
However, since the known background channels do not account for the level of the observed
background in the data sample, a scale factor is introduced to scale the MC 
background predictions for fits to the distribution in Fig.~2. The 
scale factor depends strongly on which intermediate states are considered 
for $\jpsi\ar 5 \gamma$ decays; the 
difference between the scale factors determined from different channels is treated as  
a systematic uncertainty of the background subtraction.

Figure 2 is fit with a MC-simulated $\jpsi\ar\gamma\pi^0$ histogram for the signal, a MC-simulated 
$\jpsi\ar 5 \gamma$ background shape, as well as a shape of MC simulated phase space for other 
sources of backgrounds. The number of $\gamma\pi^0$ events determined from the fit is $586\pm51$. 
The MC-determined detection efficiency for
$\jpsi\ar\gamma\pi^0,\pi^0\ar\gamma\gamma$ is 
$\varepsilon = (32.80\pm0.21)$\%, where the error comes from the limited 
statistics of the MC sample.

\begin{figure}[htbp]
\centerline{\psfig{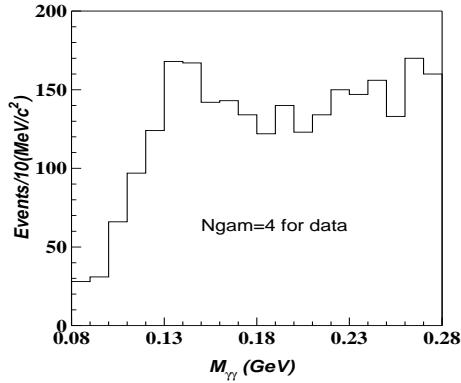}}
\caption{The invariant mass distribution of $\gamma\gamma$ pairs with the smallest opening 
angle in  $\jpsi\ar\gamma\gamma\gamma$ events selected from the four photon event sample.}
\end{figure}

\subsubsection{$\jpsi\ar\gamma\eta,\eta\ar\gamma\gamma$}
Figure 4 shows the invariant mass distribution of the two photon candidates with
the  smallest opening  angle in the $\eta$ mass region, where an $\eta$ peak is evident.

\begin{figure}[htbp]
\centerline{\psfig{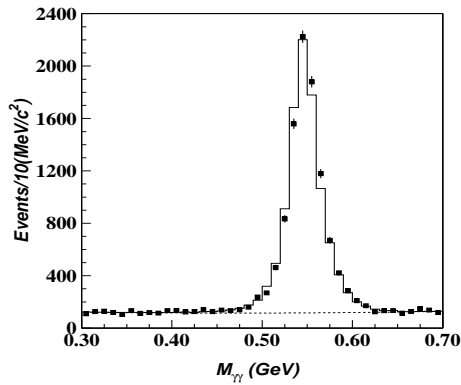}}
\caption{Invariant mass distribution of the $\gamma\gamma$ with the smallest opening 
angle of $\jpsi\ar\gamma\gamma\gamma$ candidates. Solid squares with error bars are 
data, the histogram is the fit result, and the dashed line is the background.}
\end{figure}

The $\gamma\gamma$ invariant mass distribution of Fig.~4 is fit with a 
histogram from MC-simulated
$\jpsi\ar\gamma\eta, \eta\ar\gamma\gamma$ events and a second order Legendre polynomial background
function.  The  fit yields a signal of 9096$\pm$133 $\eta$s.   The 
MC-determined detection efficiency is $\varepsilon=(36.33\pm0.22)\%$.

\subsubsection{$\jpsi\ar\gamma\etap,\etap\ar\gamma\gamma$}

Since the momentum of the $\etap$ is lower than that of the
$\pi^0$ and $\eta$ in $\jpsi$ radiative decays, the angle 
between the two $\etap$ decay photons is not small 
enough to be useful for distinguishing them from the radiative 
photon. For this channel, the mass distribution of  
the three $\gamma\gamma$ combinations for each event 
are plotted in Fig. 5, where an $\etap$ 
signal is evident above  
a smooth background due to wrong $\gamma\gamma$ combinations 
plus other background sources.

\begin{figure}[htbp]
\centerline{\psfig{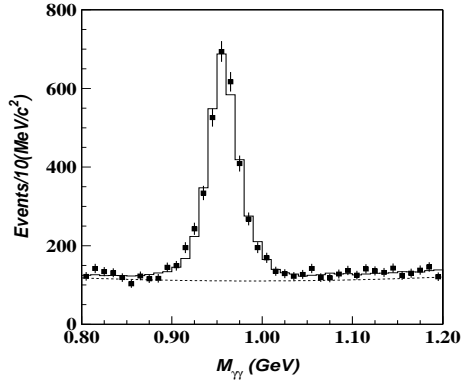}}
\caption{The $\gamma\gamma$ invariant mass distribution
for $\jpsi\ar\gamma\gamma\gamma$ candidate events (three entries per 
event).  The solid 
squares with error bars indicate data, the  histogram is the fit result, 
and the dashed line is the non-combinatorial background.}
\end{figure}

A fit to the data points, with the MC simulated mass distribution for the 
$\jpsi\ar\gamma\etap,\etap\ar\gamma\gamma$ decay including combinatorial 
background for the signal and a second order Legendre polynomial for background between 
0.8 and 1.2 $\hbox{GeV}/c^2$, yields $2982\pm101$ entries. Since all the 
$\gamma\gamma$ combinations are plotted in the $M_{\gamma\gamma}$ distribution, 
the combinatorial background is included in the entries for 
both data and MC simulation. The efficiency for signal 
$\jpsi\ar\gamma\etap,\etap\ar\gamma\gamma$ entries is
$(40.30\pm0.22)\%$. The combinatorial background is about $20\%$ for both data and MC simulation, 
and they cancel out when the $N^{obs}$ is divided by the efficiency
$40.30\%$ in the branching fraction calculation.

\subsection{\boldmath $\jpsi\ar\gamma\gamma\gamma\pi^+\pi^-$}

In this final states, there are two charged particles $\pi^+$ and $\pi^-$ 
and three photons. Candidate events are required to satisfy the following 
common selection criteria:

\begin{enumerate}
  \item Two good charged tracks with net charge zero. Each
track must have a good helix fit, a transverse momentum larger than 60 MeV/c, and 
$|\cos\theta|<0.8$, where $\theta$ is the polar angle of the track,
and  must originate from the interaction
region.
  \item At least one charged track is identified as a $\pi$, 
satisfying $\chi^2_{PID}(\pi)<\chi^2_{PID}(K)$ and 
$\chi^2_{PID}(\pi)<\chi^2_{PID}(p)$, 
where $\chi^2_{PID}=\chi^2_{dE/dx}+\chi^2_{TOF}$ is 
determined using both $dE/dx$ and TOF information.
  \item At least three photon candidates are required. The 
photon identification is similar to that used in the 
$\jpsi\ar\gamma\gamma\gamma$ analysis, except that the angle between a 
cluster and any other cluster must be greater than $18^{\circ}$, and the 
angle  between the cluster and any charged track must be greater 
than  $8^{\circ}$. These differences  reflect
different sources of fake photons.
  \item A four-constraint kinematic fit is applied to all three-photon 
combinations plus the two charged tracks assuming 
$\jpsi\ar\gamma\gamma\gamma\pi^+\pi^-$. The three-photon combination with 
the smallest $\chi^2$ is selected, and 
the $\chi^2$ of the kinematic fit is required to be less than 20. 
\end{enumerate}

The events that survive these selection criteria with an invariant mass in the range
$M_{\gamma\gamma\pi^+\pi^-}\leq1.2\hbox{GeV}/c^2$ are assumed to
come from either $\eta$ or $\etap$ decays, 
and the other photon is 
considered to be the radiative 
photon. Figure 6 shows a
scatterplot of $M_{\gamma\gamma}$ versus $M_{\gamma\gamma\pi^+\pi^-}$ for
the selected events.  Clear $\eta$ and $\etap$ signals corresponding to 
$\eta\ar\pi^0\pi^+\pi^-,\pi^0\ar\gamma\gamma$, and 
$\etap\ar\eta\pi^+\pi^-,\eta\ar\gamma\gamma$ are observed.

\begin{figure}[htbp]
\centerline{\psfig{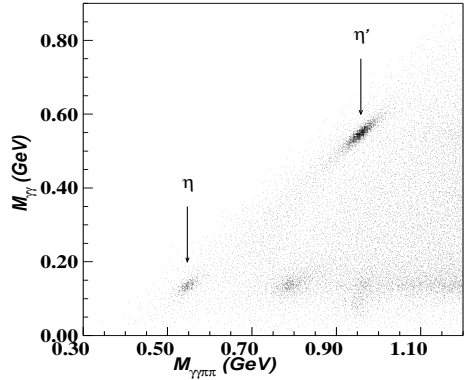}}
\caption{Scatterplot of $M_{\gamma\gamma}$ versus 
$M_{\gamma\gamma\pi^+\pi^-}$ for the 
$\jpsi\ar\gamma\gamma\gamma\pi^+\pi^-$ candidates.}
\end{figure}

\subsubsection{$\jpsi\ar\gamma\eta,\eta\ar\pi^0\pppm$}
After the requirement that the $\gamma\gamma$ invariant mass is in the
$\pi^0$ mass
region ($M_{\gamma\gamma}\in[0.088,0.182]~\hbox{GeV}/c^2$, $\pm
3 \sigma$), a clear $\eta$ signal is 
evident in the $\gamma\gamma\pi^+\pi^-$ invariant mass distribution shown 
in Fig. 7.

\begin{figure}[htbp]
\centerline{\psfig{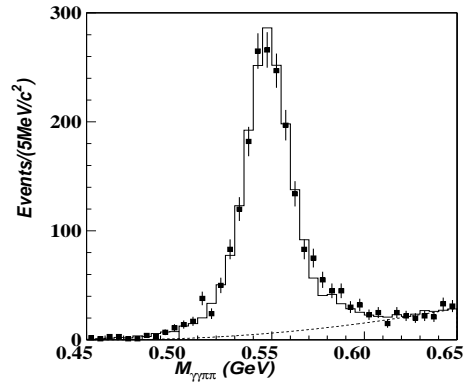}}
\caption{The $\gamma\gamma\pi^+\pi^-$ invariant mass distribution 
for $\jpsi\ar\gamma\gamma\gamma\pi^+\pi^-$ candidates that satisfy
the  requirement $M_{\gamma\gamma}\in[0.088,0.182]~\hbox{GeV}/c^2$. 
The solid squares with 
error bars indicate the data, the histogram is the fit result, and the 
dashed line is the background.}
\end{figure}

The simulated $M_{\gamma\gamma\pi^+\pi^-}$ mass distribution from the signal MC and a second-order Legendre polynomial are used to fit the 
$\gamma\gamma\pi^+\pi^-$ invariant mass distribution.
The fit gives $1885\pm58$ $\eta$ events.  The MC-determined
detection  efficiency for
$\jpsi\ar\gamma\eta,\eta\ar\pi^0\pi^+\pi^-$, and $\pi^0\ar\gamma\gamma$
is $\varepsilon=(12.25\pm0.15)\%$.

\subsubsection{$\jpsi\ar\gamma\etap,\etap\ar\eta\pppm$}

The $\gamma\gamma\pi^+\pi^-$ invariant mass distribution
for events with $\gamma\gamma$ mass within $3\sigma$ of
the $\eta$ mass ($M_{\gamma\gamma}\in[0.484,0.612]~\hbox{GeV}/c^2$),
is shown in Fig. 8.

\begin{figure}[htbp]
\centerline{\psfig{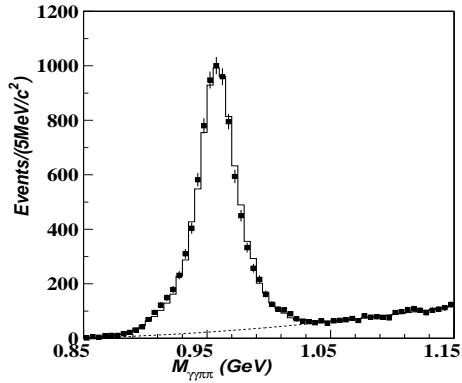}}
\caption{ The $\gamma\gamma\pi^+\pi^-$ invariant mass distribution
for events with $\gamma\gamma$ mass in the $\eta$ mass region
($M_{\gamma\gamma}\in[0.484,0.612]~\hbox{GeV}/c^2$). The solid squares 
with error bars indicate data, the histogram is the fit result, and the 
dashed  line is the background.}
\end{figure}

A similar fit as for $\eta\ar\pi^0\pi^+\pi^-$ yields $8572\pm131$ $\etap$ 
events; the MC-determined detection efficiency for
$\jpsi\ar\gamma\etap,\etap\ar\eta\pi^+\pi^-$, and $\eta\ar\gamma\gamma$ 
is $\varepsilon=(16.10\pm0.12)\%$.

\subsection{\boldmath $\jpsi\ar\gamma\gamma\pi^+\pi^-$}

$\jpsi\ar\gamma\etap$ is also studied using the $\etap\ar\gamma\pi^+\pi^-$ 
decay channel.  For this study, the $\pi^{\pm}$ and photon selection
requirements are the same as used for the 
$\jpsi\ar\gamma\gamma\gamma\pi^+\pi^-$ final state, and the event 
selection is similar, except that here at least two photons are 
required in the event. 
The photons and charged tracks are kinematically fitted to 
$\jpsi\ar\gamma\gamma\pi^+\pi^-$ assuming four-momentum conservation, and 
$\chi^2\leq20$ is required.  When there are more than two 
photons, the kinematic fit is repeated using all possible photon 
combinations, and the one with the smallest $\chi^2$ is kept. The photon with 
the higher energy is considered to be 
the radiative photon from the $\jpsi$ decay. Figure 9 shows the 
invariant mass distribution of $\gamma\pi^+\pi^-$ for the candidate 
events where an $\etap$ signal is evident.

\begin{figure}[htbp]
\centerline{\psfig{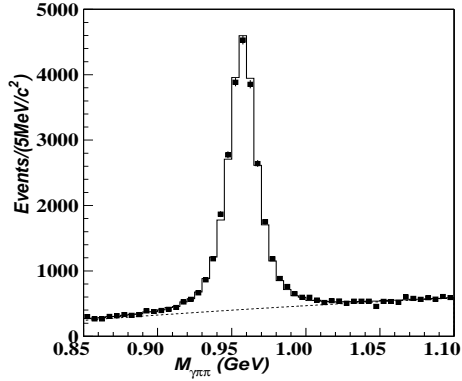}}
\caption{The $\gamma\pi^+\pi^-$ invariant mass distribution for
selected $\jpsi\ar\gamma\gamma\pi^+\pi^-$ events.  The
solid squares with error bars indicate data, the histogram is the fit 
result, and the dashed line is the background.}
\end{figure}

Figure 9 shows the result of a fit to the $\gamma\pi^+\pi^-$ invariant 
mass  distribution that follows a similar procedure as that for the 
fit to the $\gamma\gamma\pi^+\pi^-$ distribution of the previous section.
The fit yields  $23243\pm229$ $\etap$ signal events.  
The MC-determined detection efficiency for 
$\jpsi\ar\gamma\etap,\etap\ar\gamma\pi^+\pi^-$ 
is $\varepsilon=(25.02\pm0.10)\%$.

\section{SYSTEMATIC ERRORS}
Systematic errors in the branching fraction
measurements mainly originate 
from photon identification (ID), MDC tracking efficiency, particle ID, 
kinematic fitting, mass resolution, $\pi^0$ reconstruction, 
and parameterizations of background 
shapes.

\subsection{Photon identification}

The efficiency for photon ID is discussed in Ref.~\cite{lism}. It 
is found that the
relative  efficiency difference between data and MC simulation
for high energy photon detection is about 0.8\% per photon, 
while for low energy photons, the difference is around 
2\% per photon. Since the energy of the radiative photon 
in  $\jpsi\ar\gamma\pi^0,\gamma\eta$, and $\gamma\etap$ is high, 
and the energies of the photons from pseudoscalar particle decays are low, 
the total systematic error due to photon ID is taken as $(0.8+2.0n)\%$,
where $n$ is the number of photons from the 
pseudoscalar particle decay.

\subsection{MDC tracking}
The MDC tracking efficiency is studied in Ref.~\cite{simbes}. It is found 
that there is a 2.0\% relative difference per track between data and MC 
simulation. For the channels in this analysis that have two charged 
tracks, a 4\% systematic error on the MDC tracking efficiency is
assigned.

\subsection{Particle ID}
A clean charged $\pi$ sample obtained from $\jpsi\ar\rho\pi$ without 
the use of particle ID is used to study data-MC differences between
particle ID efficiencies for
different momentum ranges. Since only one 
of the two charged tracks is required to be identified as a pion, the MC 
simulates data rather well; 
it is found that the MC simulation agrees 
with data within 0.2\% for both $\jpsi\ar\gamma\gamma\gamma\pi^+\pi^-$ 
and $\jpsi\ar\gamma\gamma\pi^+\pi^-$ modes.

\subsection{Kinematic fit}

Samples of $\jpsi\ar\rho\pi$ and $e^+e^-\ar\gamma\gamma$ 
events selected without using kinematic fits
are used to study the systematic error associated with the
four-constraint kinematic fit.  For the  $\chi^2\leq20$ criteria, 
the difference of kinematic fit efficiencies between data and MC
simulation is less than 1.2\% for $\rho\pi$, and
2.4\% for $e^+e^-\ar\gamma\gamma$.  Extrapolating these differences to the
channels reported here, we conservatively assign a 4\%  systematic 
error to the kinematic fit efficiency.

\subsection{Different mass resolution between MC and DATA}

There is a slight difference of the mass resolution between MC
simulation and data. When the histogram shape of invariant mass
distribution from MC simulation is used to fit the invariant mass
distribution of data, it introduces some systematic error.
The high statistics decay channels
$\jpsi\ar\gamma\eta,\eta\ar\gamma\gamma$,
$\jpsi\ar\gamma\etap,\etap\ar\gamma\gamma\pi^+\pi^-$ and
$\jpsi\ar\gamma\etap,\etap\ar\gamma\pi^+\pi^-$ are used to study this
source of systematic error.  For these channels,
we allow the mass resolution to vary in the fit to the
invariant mass distributions, and we also determine the number of 
signal events by subtracting side-band-estimated backgrounds.
The resulting
branching fractions change by at most 1.6\%, 0.1\%, and 0.6\% for
$\jpsi\ar\gamma\eta,\eta\ar\gamma\gamma$,
$\jpsi\ar\gamma\etap,\etap\ar\eta\pi^+\pi^-$, and
$\jpsi\ar\gamma\etap,\etap\ar\gamma\pi^+\pi^-$ respectively. 
We assign 1.6\%,
0.1\% and 0.6\%  as the systematic errors due to 
mass resolution uncertainties for the $\jpsi\ar\gamma\gamma\gamma$,
$\jpsi\ar\gamma\gamma\gamma\pi^+\pi^-$ and
$\jpsi\ar\gamma\gamma\pi^+\pi^-$ decay modes, respectively.

\subsection{\boldmath Reconstruction of $\pi^0$}
In $\jpsi\ar\gamma\pi^0$, the $\pi^0$ momentum is high and 
the angle between the two decay photons is small. As a result, it
is possible for the two  photons to merge into a single BSC cluster. 
According to a study reported in 
Ref.~\cite{xinbo}, the systematic error associated with 
1.5~GeV $\pi^0$ reconstruction is 0.83\%. 
The effect on low energy $\pi^0$s 
or $\eta$s is small enough to be neglected.

\subsection{Background shape}
For the $\jpsi\ar\gamma\pi^0$ mode, the background estimate based on the
four photon event sample has a large uncertainty.  Fits using
MC-determined background shapes from different background channels
yield different numbers of signal events; 
the corresponding changes in the
branching fractions range between $^{+16.4}_{-6.8}\%$. 
The largest difference
is taken as the systematic error.  
Different order Legendre polynomials
are used to fit the mass spectra for the other decay modes, and the
differences between these fits and those used to get the numbers of
signal events are used as the systematic error due to background 
parameterization. Different fitting ranges are also used in the fit, and 
the differences are included in the systematic error. The uncertainty due 
to the background shape and fitting range is less than 2\%.

\subsection{Branching fractions of the secondary decays}
The branching fractions of decay from $\pi^{0},\eta$ and $\eta^{\prime}$ 
are taken from the PDG~\cite{PDG04}; the uncertainties are included in 
the measurement errors of the reported branching fractions.

\subsection{\boldmath The number of $J/\psi$ events}
The total number of $J/\psi$ events, determined from the 4-prong data sample,
is $(57.7\pm2.72)\times10^{6}$.
 The 4.72\% relative error is taken 
as a systematic error~\cite{fangss}.

\subsection{Total systematic error}

Table I summarizes the systematic errors from all sources for each 
mode.  We assume all the sources are independent and
add them in quadrature; the resulting total systematic errors are 
$^{+18.3}_{-10.6}\%$, $8.1\%$, $10.6\%$, $9.3\%$, $9.5\%$, and $8.7\%$ 
for $\jpsi\ar\gamma\pi^0\ar\gamma\gamma\gamma$, 
$\jpsi\ar\gamma\eta\ar\gamma\gamma\gamma$, 
$\jpsi\ar\gamma\etap\ar\gamma\gamma\gamma$, 
$\jpsi\ar\gamma\eta\ar\gamma\gamma\gamma\pi^+\pi^-$, 
$\jpsi\ar\gamma\etap\ar\gamma\gamma\gamma\pi^+\pi^-$, and 
$\jpsi\ar\gamma\etap\ar\gamma\gamma\pi^+\pi^-$, respectively.

\begin{table}[htbp]
\begin{center}
\caption{Summary of the systematic errors (\%). } \vspace{3mm}
  \begin{tabular}{lcccccc} \hline
Sources &$\pi^{0}\ar\gamma\gamma$ &$\eta\ar\gamma\gamma$&$\eta^{\prime}\ar\gamma\gamma$ & $\eta\ar\pi^{0}\pi^{+}\pi^{-}$&$\eta^{\prime}\ar\eta\pi^{+}\pi^{-}$&$\eta^{\prime}\ar\gamma\pi^+\pi^-$ \\
\hline
Photon ID & 4.8&4.8&4.8&4.8&4.8&2.8\\
Tracking  & - &-&-&4.0&4.0&4.0\\
Particle ID & -&-&-&0.2&0.2&0.2\\
Kinematic fit & 4.0&4.0&4.0&4.0&4.0&4.0\\
Mass resolution & 1.6&1.6&1.6&0.1&0.1&0.6\\ 
$\pi^0$ reconstruction & 0.83 &-&- &- &- &- \\
Background shape & $^{+16.4}_{-6.8}$ &0.73 &1.8 &1.7 &0.5 &0.2 \\
Branching fraction used & 0.04&0.66&6.61&1.77&3.45&3.39\\
Number of $J/\psi$ & 4.72&4.72&4.72&4.72&4.72&4.72\\
Statistic of MC sample & 0.64 & 0.61& 0.55& 1.23& 0.75& 0.40 \\
\hline
Total error & $^{+18.3}_{-10.6}$ & 8.1 & 10.6 & 9.3 & 9.5 & 8.7 \\
\hline
\end{tabular}

\end{center}
\end{table}

\section{RESULTS AND DISCUSSION}

The branching fractions of $\jpsi$ decays are determined from the relation
$$
Br(\jpsi\ar\gamma P)=\frac{N^{obs}(\jpsi\ar\gamma P \ar \gamma 
Y)}{N^{\jpsi}\cdot Br(P \ar Y) \cdot \varepsilon(\jpsi\ar\gamma P 
\ar\gamma Y)},
$$
where $P$ is either $\pi^0$, $\eta$, or $\etap$, $Y$ is the pseudoscalar 
decay final state, and $Br(P \ar Y)$ is the branching fraction of the 
pseudoscalar decays into final state $Y$. The results of $Br(\jpsi\ar\gamma P)$ are listed in Table~II.


The branching fractions of 
$\jpsi\ar\gamma\eta$, $\jpsi\ar\gamma\etap$ measured from different decay 
modes are consistent with each other within the statistical 
fluctuations and uncommon systematic errors.  The measurements from 
the different modes are, therefore, combined using a standard weighted 
least-squares procedure taking into consideration the correlations 
between the 
measurements; the mean value and the error are calculated by:
$$
\bar{x}\pm\delta\bar{x}=\frac{\sum_{j}{x_{j}}\cdot(\sum_{i}{\omega_{ij}})}{\sum_{i}\sum_{j}{\omega_{ij}}}\pm\sqrt{\frac{1}{\sum_{i}\sum_{j}{\omega_{ij}}}}.
$$
Here $\omega_{ij}$ is the element of the 
weighted matrix $W=V^{-1}_x$, where $V_x$ is the covariance matrix calculated according 
to the systematic errors listed in Table I. For $\jpsi\ar\gamma\eta$, the correlation 
coefficient between $\eta\ar\gamma\gamma$ and $\eta\ar\gamma\gamma\pi^+\pi^-$ 
is $\rho(1,2)=0.553$; for $\jpsi\ar\gamma\etap$, the correlation coefficients 
between $\etap\ar\gamma\gamma$, $\etap\ar\gamma\pi^+\pi^-$ 
and $\etap\ar\gamma\gamma\pi^+\pi^-$ are $\rho(1,2)=0.296$, $\rho(1,3)=0.404$ 
and $\rho(2,3)=0.703$. The weighted averages of BESII measurements and the 
PDG~\cite{PDG04} values are listed in Table II.

\begin{table}
\caption{Branching fractions of
$J/\psi\rightarrow\gamma\pi^{0}, \gamma\eta$ and $\gamma\eta^{\prime}$.}
\begin{center}
\footnotesize
  \begin{tabular}{|c|l|r|r|c|} 
\hline
 \multicolumn{2}{|c|}{Decay mode}   &   BESII    & BESII combined & PDG~\cite{PDG04}  \\
\hline
 $\gamma\pi^0$ & $\pi^{0}\rightarrow\gamma\gamma$ & {\footnotesize$(3.13\pm0.28^{+0.58}_{-0.34})\times10^{-5}$ }& $(3.13^{+0.65}_{-0.44})\times10^{-5}$ & $(3.9\pm1.3)\times10^{-5}$\\
\hline
$\gamma\eta$ &$\eta\rightarrow\gamma\gamma$  & {\footnotesize $(11.00\pm0.16\pm0.90)\times10^{-4}$}&$(11.23\pm0.89)\times10^{-4}$ &$(8.6\pm0.8)\times10^{-4}$\\
\cline{2-3}
 & $\eta\rightarrow\pi^{0}\pi^{+}\pi^{-}$  &{\footnotesize$(11.94\pm0.37\pm1.11)\times10^{-4}$}& &\\
\hline
 &$\eta^{\prime}\rightarrow\gamma\gamma$  &{\footnotesize$(6.05\pm0.21\pm0.65)\times10^{-3}$}&&\\
\cline{2-3}
 $\gamma\eta^{\prime}$ & $\eta^{\prime}\rightarrow\gamma\rho$  &{\footnotesize$(5.46\pm0.06\pm0.48)\times10^{-3}$} & $(5.55\pm0.44)\times10^{-3}$ &$(4.31\pm0.30)\times10^{-3}$\\
\cline{2-3}
 &$\eta^{\prime}\rightarrow\eta\pi^{+}\pi^{-}$  &{\footnotesize$(5.28\pm0.08\pm0.51)\times10^{-3}$}&&\\
\hline
  \end{tabular}
\end{center}
\end{table}

\begin{figure}[htbp]
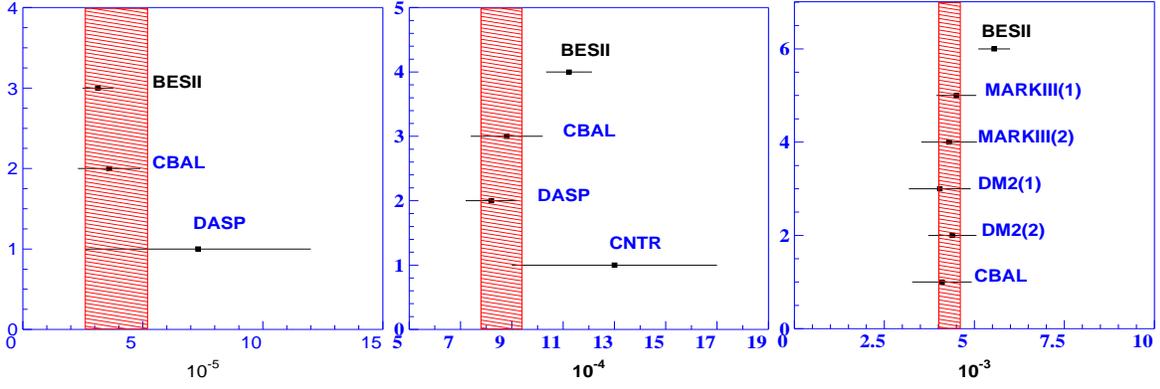

\centerline{\psfig{file=10a.epsi,width=5.0cm,height=5.0cm}
            \psfig{file=10b.epsi,width=5.0cm,height=5.0cm}
            \psfig{file=10c.epsi,width=5.0cm,height=5.0cm}}
\caption{Comparisons of $Br(\jpsi\ar\gamma\pi^0)$,
  $Br(\jpsi\ar\gamma\eta)$ and $Br(\jpsi\ar\gamma\etap)$ between BESII
  and previous measurements~\cite{PDG04}. The shaded regions are the
  world averages from the PDG~\cite{PDG04}.}
\end{figure}

Figure 10 shows the comparisons between the measurements in this paper
and those from previous measurements~\cite{DESY,CBAL,MARK3,DM2,DASP}.
Our measurement of $Br(\jpsi\ar\gamma\pi^0)$ agrees with those of
Crystal Ball~\cite{CBAL} and DASP~\cite{DASP} within the large errors 
of the previous measurements, and has much improved precision. Our
measurement's lower central value may be because background channels
that produce a peak in the signal region have been considered. Our 
measurements of $Br(\jpsi\ar\gamma\eta)$ and $Br(\jpsi\ar\gamma\etap)$ are 
higher than the PDG world average~\cite{PDG04}, and have better precision
than the previous measurements~\cite{DESY,CBAL,MARK3,DM2}.

The results listed in Table II also allow us calculate the relative branching 
fractions for $\eta$ and $\etap$ decays; considering the common errors in the 
measurements, one gets
{\footnotesize
$$
\frac{Br(\etap\ar\gamma\gamma)}{Br(\etap\ar\gamma\pi^+\pi^-)}=\frac{Br(\jpsi\ar\gamma\etap,\etap\ar\gamma\gamma)}{Br(\jpsi\ar\gamma\etap,\etap\ar\gamma\pi^+\pi^-)}=0.080\pm0.008,
$$
$$
\frac{Br(\etap\ar\eta\pi^+\pi^-)}{Br(\etap\ar\gamma\pi^+\pi^-)}=\frac{Br(\jpsi\ar\gamma\etap,\etap\ar\eta\pi^+\pi^-)}{Br(\jpsi\ar\gamma\etap,\etap\ar\gamma\pi^+\pi^-)}=1.45\pm0.07,
$$
$$
\frac{Br(\eta\ar\gamma\gamma)}{Br(\eta\ar\pi^0\pi^+\pi^-)}=\frac{Br(\jpsi\ar\gamma\eta,\eta\ar\gamma\gamma)}{Br(\jpsi\ar\gamma\eta,\eta\ar\pi^0\pi^+\pi^-)}=1.61\pm0.14.
$$}
The correlation coefficients between denominator and numerator in above 
equations are 0.419, 0.859 and 0.575 respectively. 
The world averages~\cite{PDG04} of the same ratios are $0.072\pm0.006$, $1.50\pm0.08$ 
and $1.75\pm0.04$ respectively. The agreement is quite good.

If both the OZI rule and the $SU_f(3)$ symmetry are exact, it is expected that~\cite{thetap1}:
$$
R=\frac{\Gamma(\jpsi\ar\gamma\etap)}{\Gamma(\jpsi\ar\gamma\eta)}=(\frac{P_{\etap}}{P_{\eta}})^3\cdot{\cot^2\theta_P}.
$$

Using $Br(\jpsi\ar\gamma\eta)$ and $Br(\jpsi\ar\gamma\etap)$ in this analysis, one obtains
$$
R=4.94\pm0.40,
$$
$$
|\theta_P|=(22.08\pm0.81)^{\circ},
$$
where the common errors have been considered in the ratio calculation. 
Comparing with the mixing models with states other than $\eta$ and $\etap$, 
the measurement of $R$ agrees with the prediction of $R=5.1$~\cite{ratio2} within 
one standard deviation, while it deviates from $R=3.9$~\cite{ratio1} by more than 3 standard 
deviations. According to the theoretical calculation of Ref.~\cite{mixing}, the value of 
$\theta_P$ is negative, in which case its value is $\theta_P=(-22.08\pm0.81)^{\circ}$. 

\section{SUMMARY}
Using 58 million $\jpsi$ events collected by BESII, 
the branching fractions of $\jpsi$ decays into a photon 
and a pseudoscalar meson are measured as 
$Br(\jpsi\ar\gamma\pi^0)=(3.13^{+0.65}_{-0.44})\times 10^{-5}$, 
$Br(\jpsi\ar\gamma\eta)=(11.23\pm 0.89)\times 10^{-4}$, and 
$Br(\jpsi\ar\gamma\etap)=(5.55\pm 0.44)\times 10^{-3}$. 
The results are compared to $\eta$ and $\etap$ mixing models.

\section{Acknowledgment}
The BES collaboration thanks the staff of BEPC for their hard
efforts. This work is supported in part by the National Natural
Science Foundation of China under contracts Nos. 10491300,
10225524, 10225525, 10425523, the Chinese Academy of Sciences under
contract No. KJ 95T-03, the 100 Talents Program of CAS under
Contract Nos. U-11, U-24, U-25, and the Knowledge Innovation
Project of CAS under Contract Nos. U-602, U-34 (IHEP), the
National Natural Science Foundation of China under Contract No.
10225522 (Tsinghua University), and the Department of Energy under
Contract No.DE-FG02-04ER41291 (U Hawaii).



\begin{thebibliography}{dd}
\bibitem{theo} J. F. Donoghue, B. R. Holstein and Y. -C. R. Lin, Phys.
Rev. Lett. {\bf 55}, 2766 (1985); R. Escribano and J. -M. Fr$\grave{e}$re, hep-ph/0501072 (2005).
\bibitem{mixing} F. J. Gilman and R. Kauffman, Phys. Rev. {\bf D36}, 2761 (1987).
\bibitem{thetap1} R. N. Cahn and M. S. Chanowitz, Phys. Lett. {\bf B59}, 277 (1975).
\bibitem{pertu1} J. G. K\"{o}rner, J. H. K\"{u}hn, M. Krammer, H. Schneider, Nucl. Phys. {\bf B229}, 115 (1983).
\bibitem{pertu2} B. Guberina and J. H. K\"{u}hn, Nuovo Cim. Lett. {\bf 32}, 295 (1981).
\bibitem{PDG04}  S. Eidelman {\it et al.}(Particle Data Group), Phys. Lett. {\bf B592}, 1 (2004).
\bibitem{ratio1} H. Fritsch and J. D. Jackson, Phys. Lett. {\bf B66}, 365 (1977).
\bibitem{ratio2} H. Yu, High Energy Phys. \& Nucl. Phys. {\bf 12}, 754 (1988) (in Chinese).
\bibitem{DESY} W. Bartel {\it et al.}, Phys. Lett. {\bf B66}, 489 (1977).
\bibitem{CBAL} E. D. Bloom, C. Peck, ARNS {\bf 33}, 143 (1983).
\bibitem{MARK3} T. Bolton {\it et al.}, Phys. Rev. Lett. {\bf 69}, 1328 (1992).
\bibitem{DM2} J. E. Augustin {\it et al.}, Phys. Rev. {\bf D42}, 10 (1990).
\bibitem{DASP} W. Braunschweig {\it et al.}, Phys. Lett. {\bf B67}, 243 (1977).
\bibitem{VMD} V. L. Chernyak and A. R. Zhitnitsky, Phys. Rept. {\bf 112}, 173 (1984).
\bibitem{mutipole} Y. P. Kuang, Phys. Rev. {\bf D42}, 2300 (1990).
\bibitem{bes2} BES Collab., J. Z. Bai {\it et al.}, Nucl. Instrum. Methods {\bf A458}, 627 (2001).
\bibitem{simbes} BES Collab., M. Ablikim {\it et al.}, physics/0503001, to be published in Nucl. Instrum. Methods A.
\bibitem{lism} S. M. Li {\it et al.}, High Energy Phys. \& Nucl.Phys., {\bf 28}, 859 (2004) (Chinese Edition).
\bibitem{xinbo} BES Collab., M. Ablikim {\em et al.}, Phys. Rev. {\bf
D71}, 072006 (2005).
\bibitem{fangss} S. S. Fang {\it et al.},
High Energy Phys. \& Nucl. Phys. {\bf 27}, 277 (2003) (in Chinese).

\end{thebibliography}
\end{document}